\documentclass[pra,twocolumn,showpacs,letterpaper,showpacs,superscriptaddress,nofootinbib]{revtex4-1}
\usepackage{graphicx,amsmath,amssymb,amsfonts,latexsym,color,dcolumn,bm,subfigure}

\newcommand{\p}{{\rm p}}
\newcommand{\s}{{\rm S}}
\newcommand{\B}{{\rm B}}
\newcommand{\ba}{\begin{align}}
\newcommand{\eA}{\end{align}}

\begin{document}

\title{Long-lived guided phonons in fiber by manipulating two-level systems}

\author{R. O. Behunin}
\affiliation{Department of Applied Physics, Yale University, New Haven, Connecticut 06511, USA}
\author{P. Kharel}
\affiliation{Department of Applied Physics, Yale University, New Haven, Connecticut 06511, USA}
\author{W. H. Renninger}
\affiliation{Department of Applied Physics, Yale University, New Haven, Connecticut 06511, USA}
\author{H. Shin}
\affiliation{Department of Applied Physics, Yale University, New Haven, Connecticut 06511, USA}
\author{F. W. Carter}
\affiliation{Department of Applied Physics, Yale University, New Haven, Connecticut 06511, USA}
\author{E. Kittlaus}
\affiliation{Department of Applied Physics, Yale University, New Haven, Connecticut 06511, USA}
\author{P. T. Rakich}
\affiliation{Department of Applied Physics, Yale University, New Haven, Connecticut 06511, USA}

\date{ \today}

\begin{abstract}
The synthesis of ultra-long lived acoustic phonons in a variety of materials and device geometries could enable a range of new coherent information processing and sensing technologies; many forms of phonon dissipation pose a barrier to this goal. We explore linear and nonlinear contributions to phonon dissipation in silica at cryogenic temperatures using fiber-optic structures that tightly confine both photons and phonons to the fiber-optic core.  When immersed in helium, this fiber system supports nearly perfect guidance of 9 GHz acoustic phonons; strong electrostrictively mediated photon-phonon coupling (or guided-wave stimulated Brillouin scattering) permits a flexible form of laser-based phonon spectroscopy. 
Through linear and nonlinear phonon spectroscopy, we isolate the effects of disorder-induced two-level tunneling states as a source of phononic dissipation in this system. We show that an ensemble of such two-level tunneling states can be driven into transparency--virtually eliminating this source of phonon dissipation over a broad range of frequencies. Experimental studies of phononic self-frequency saturation show excellent agreement with a theoretical model accounting for the phonon coupling to an ensemble of two-level tunneling states. Extending these results, we demonstrate a general approach to suppress dissipation produced by two-level tunneling states via cross-saturation,  where the lifetime of a phonons at one frequency can be extended by the presence of a high intensity acoustic beam at another frequency. Our modeling and measurements suggest that Rayleigh scattering dominates phonon losses for the longest lifetimes achieved in our system.  Although these studies were carried out in silica, our findings are quite general, and can be applied to a range of materials systems and device geometries. 
\end{abstract}


\maketitle


\section{Introduction}

Access to new regimes of classical and quantum dynamics hinge upon our ability to create and manipulate ultra-long lived coherent excitations in electromagnetic, optical, and phononic domains 
\cite{Martinis02,OptomechanicsBook,Galliou13,Goryachev13c,Aspelmeyer13,Andrews14,Bagci14,Bochmann13,Pitanti14,Goryachev13}. 
In particular, ultra long-lived phonon modes have been identified as a crucial new resource by quantum information, optomechanics, and precision metrology communities 
\cite{O'Connell08,O'Connell10,Regal08,Teufel09,Anetsberger09,Westphal12,Kippenberg07,Kippenberg08,Marquardt09,Aspelmeyer13,Galliou13,Goryachev13c}. 
To this end, a variety of systems, ranging from nano- and micro-scale phononic devices to resonator technologies of centimeter-scale, have harnessed remarkable phonon coherence times \cite{Galliou13,Aspelmeyer13,Meenehan14,Seoanez07,Sankey10,O'Connell10,OptomechanicsBook,Shin13}; however, radical improvements in performance are possible if technical and fundamental sources of dissipation are mastered 
\cite{Galliou13,Aspelmeyer13,Goryachev13,Seoanez07,Goryachev13b,Goryachev12}. 
This realization has spawned a resurgence of interest in the fundamental origins of phonon dissipation at cryogenic temperatures \cite{Arcizet09,Riviere11,Meenehan14,Galliou11,Goryachev12,Goryachev13b,Faust14}.  

A ubiquitous source of dissipation arises from disorder-induced defects. Some of such defects possess quantized energy spectra, and can exchange energy with electromagnetic, optical, and phononic fields 
\cite{Martinis05,Constantin09,MacQuarrie13,MacQuarrie14,Neeley08,Seoanez07,Meenehan14,Grabovskij12,Golding73,Golding76,Golding76b}. 
When phonon-active, such defect states can absorb and emit phonons just as atoms absorb and emit light. 
Dissipation by such defects pose a fundamental limit to phonon lifetimes in acoustic media at cryogenic temperatures. Phonon-active defect states have been extensively studied in (highly disordered) amorphous media \cite{Anderson72,Jackle72,Hunklinger82,Phillips87,Golding73,Golding76,Golding76b,Arcizet09,Riviere11,Faust14}; however, their deleterious effects also appear within highly ordered crystalline systems and in systems with material interfaces 
\cite{Galliou13,Goryachev13,Meenehan14}. This form of dissipation bars access to new regimes of classical and quantum dynamics, central to a range of emerging technologies \cite{Galliou13,Goryachev13}.

In this work, we examine the dynamics of an ensemble of phonon-active defects using a guided wave geometry that produces tight confinement of both light and acoustic waves (Fig 1). Strong photon-phonon coupling within this system permits noninvasive optical excitation and interrogation of high frequency ($\sim 9.2$ GHz) phonons.  Tight confinement of acoustic modes produce high phonon intensities ($600$ W/m$^2$) in the core of this waveguide, permitting frequency selective nonlinear phonon spectroscopy with modest ($\sim$mW) optical powers over a range of temperatures (1.1-300 K). Building on established models, we elucidate the nature of phonon-defect interactions in our system, allowing us to extract defect density, coupling strength, and a range of other parameters that capture the dynamics of the defect ensemble. We show that this ensemble of defects can be driven into transparency in the strong-field limit yielding an estimated factor 45  suppression of defect-induced dissipation at 1.1 Kelvin. In this limit, defect-induced dissipation is a negligible source of loss within our guided-wave system; remarkably high phononic Q-factors ($>$12,000) and decay lengths ($>$1mm) are achieved. Building on these findings, material engineering and more sophisticated schemes could provide a path toward radically enhanced coherence times \cite{Krimer15} in a range of systems, as the basis for emerging quantum information technologies.

Phonon dissipation can be suppressed by lowering the
system temperature (below 150K in our system), and
may in itself be a viable strategy to achieve low phonon
losses in crystalline material which have low defect densities. However, for amorphous (and even crystalline) materials such as silica or silicon nitride, a temperature will be reached below which the acoustic damping will cease to decrease as the temperature is lowered further \cite{Pohl02,LeFloch03,Sonehara07,Galliou13,Goryachev13,Goryachev13c}.
This effect is believed to arise from resonant absorption
by hypothetical two-level tunneling states (TLSs) \cite{Anderson72,Jackle72,Hunklinger82,Phillips87};
the dominant contribution to phonon dissipation in many
materials at low temperatures.
 
\begin{figure*}[t!]
\begin{center}
\includegraphics[width=\textwidth]{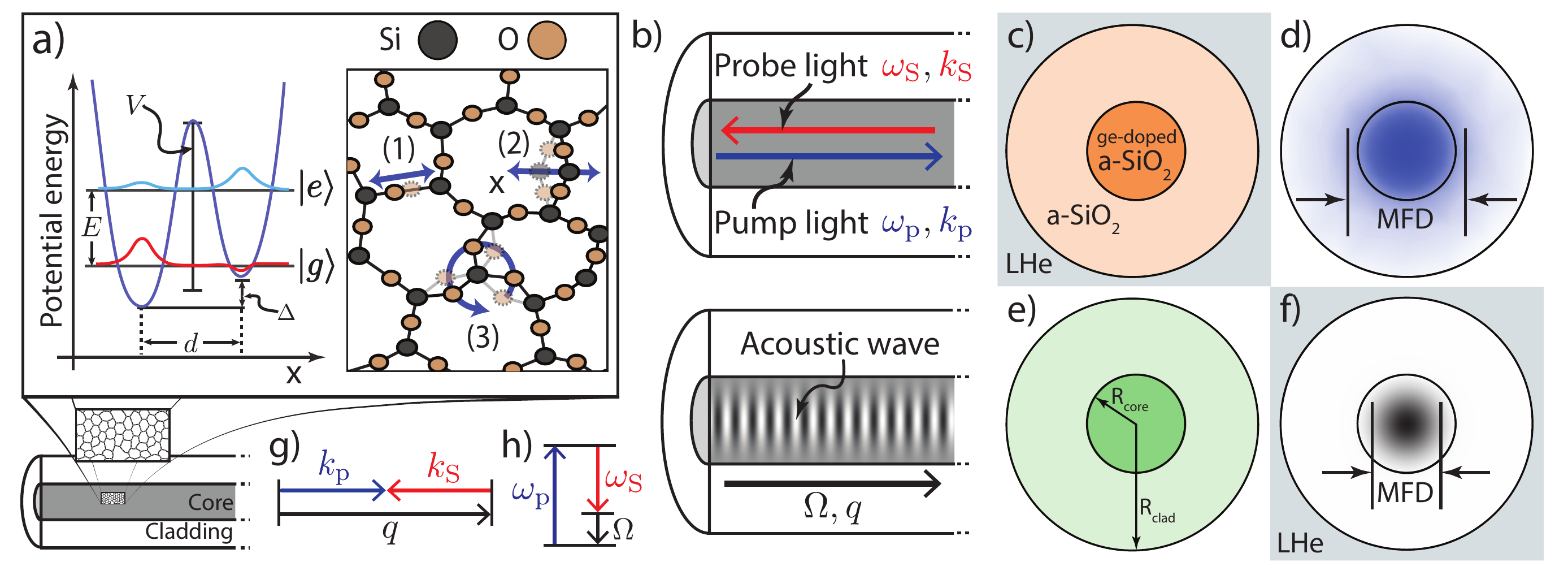}
\caption{a) Illustration of the microscopic structure of silica glass and some candidate defects proposed as the origin of two-level tunneling states (1)-(3). 
Individual TLSs are characterized by an asymmetric double-well potential for the generalized coordinate x with 
well-separation $d$, asymmetry $\Delta$, barrier height $V$, and the average oscillation frequency of the two individual wells  $\omega_c$. Tunneling between the two-wells is characterized by the tunneling strength $\Delta_0$ (defined below).  Excited $|e \rangle$ and ground $|g \rangle$ energy eigenstates of an uncoupled TLS are split by energy $E = \sqrt{\Delta^2 +\Delta^2_0}$. 
b) Guided optical and acoustic beams used for the stimulated Brillouin spectroscopy of our system.
c) \& e) define the material properties and geometry of the fiber system under study. LHe stands for liquid helium. 
Optical d) and acoustic f) intensity profile for the modes participating in Brillouin scattering. MFD is mode field diameter. 
g) spatial and h) temporal phase matching conditions between the optical and acoustic waves in SBS (see text). 
}
\label{TunnelingStates}
\end{center}
\end{figure*}

\begin{figure}
\begin{center}
\includegraphics[width=3.3in]{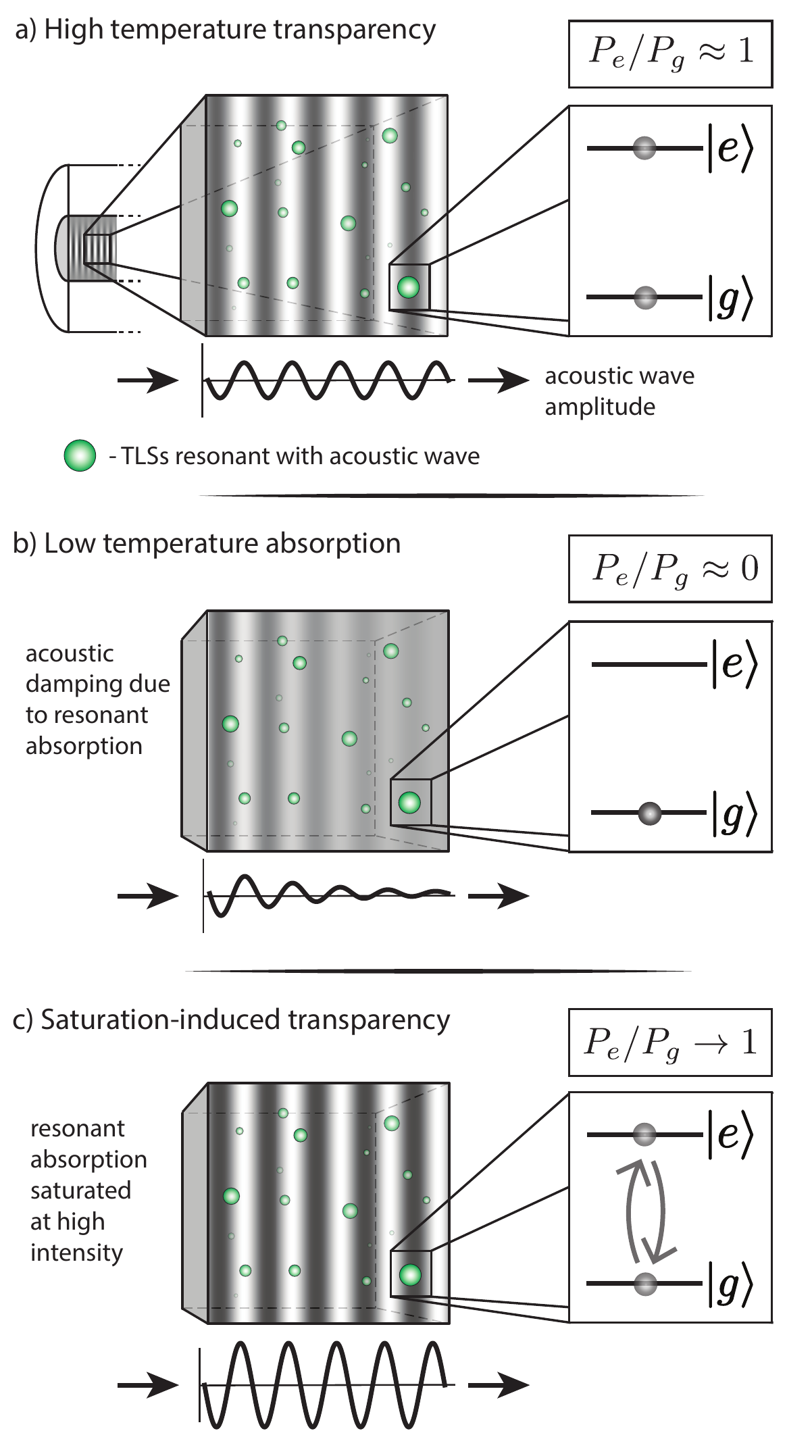}
\caption{Resonant absorption of phonons of angular frequency $\Omega$ by TLSs (illustrated by green spheres). 
a) At high temperatures ($k_B T \gg \hbar \Omega$) the TLSs addressed by the acoustic wave have almost equal probabilities of being in the excited or ground states, and hence resonant absorption is compensated by stimulated phonon emission. 
b) At low temperatures ($k_B T \ll \hbar \Omega$) the TLSs addressed by the acoustic wave condense to their ground states.  
At sufficiently low acoustic intensities TLSs, having resonantly absorbed a phonon, decay via spontaneous emission thus attenuating the incident acoustic wave in the process. 
c) At low temperatures and high phonon intensities a given TLS is interrogated by multiple phonons during its excited state lifetime. Hence, an excited TLS may decay via stimulated phonon emission which preserves the coherence of the incident phonon. At high intensities resonant absorption is compensated by stimulated emission. 
}
\label{TunnelingStates-2}
\end{center}
\end{figure}

Tunneling states are hypothesized to arise from a subset of atoms that inhabit asymmetric double-well potentials  \cite{Anderson72,Jackle72,Hunklinger82,Phillips87} (see Fig.\ref{TunnelingStates}). A perturbation of this potential by strain enables phonon absorption, and at low temperatures when these TLSs condense into their ground states (see Figs. \ref{TunnelingStates} and \ref{TunnelingStates-2}) they are capable of resonant absorption of phonons with energy $E$. The two-level nature of this decay channel leads to non-linearity in the phonon dynamics \cite{Anderson72,Jackle72,Hunklinger82,Phillips87,Golding73} and provides a way to break through the dissipation floor established by resonant absorption. 

We demonstrate that by working with high phonon intensities resonant absorption can be saturated, significantly extending phonon lifetimes. This self-frequency saturation is accomplished by increasing the amplitude of an acoustic mode (at a single frequency) to drive the TLSs into transparency. 

Two-level tunneling states have generated much contemporary interest in their own right for the possibility of beneficial implications. In particular, it has been shown that they can engender the dynamics of nano-electromechanical systems with nonlinearity, even in the single phonon regime \cite{Ramos13}, and their relatively long coherence times have facilitated their use as a quantum memory \cite{Neeley08}. In light of this recent interest, our exploration of TLSs and their manipulation may provide valuable information for a variety fields. 

\section{Overview} 

\subsection{Stimulated Brillouin scattering}

\begin{figure}[h]
\begin{center}
\includegraphics[width=3.3in]{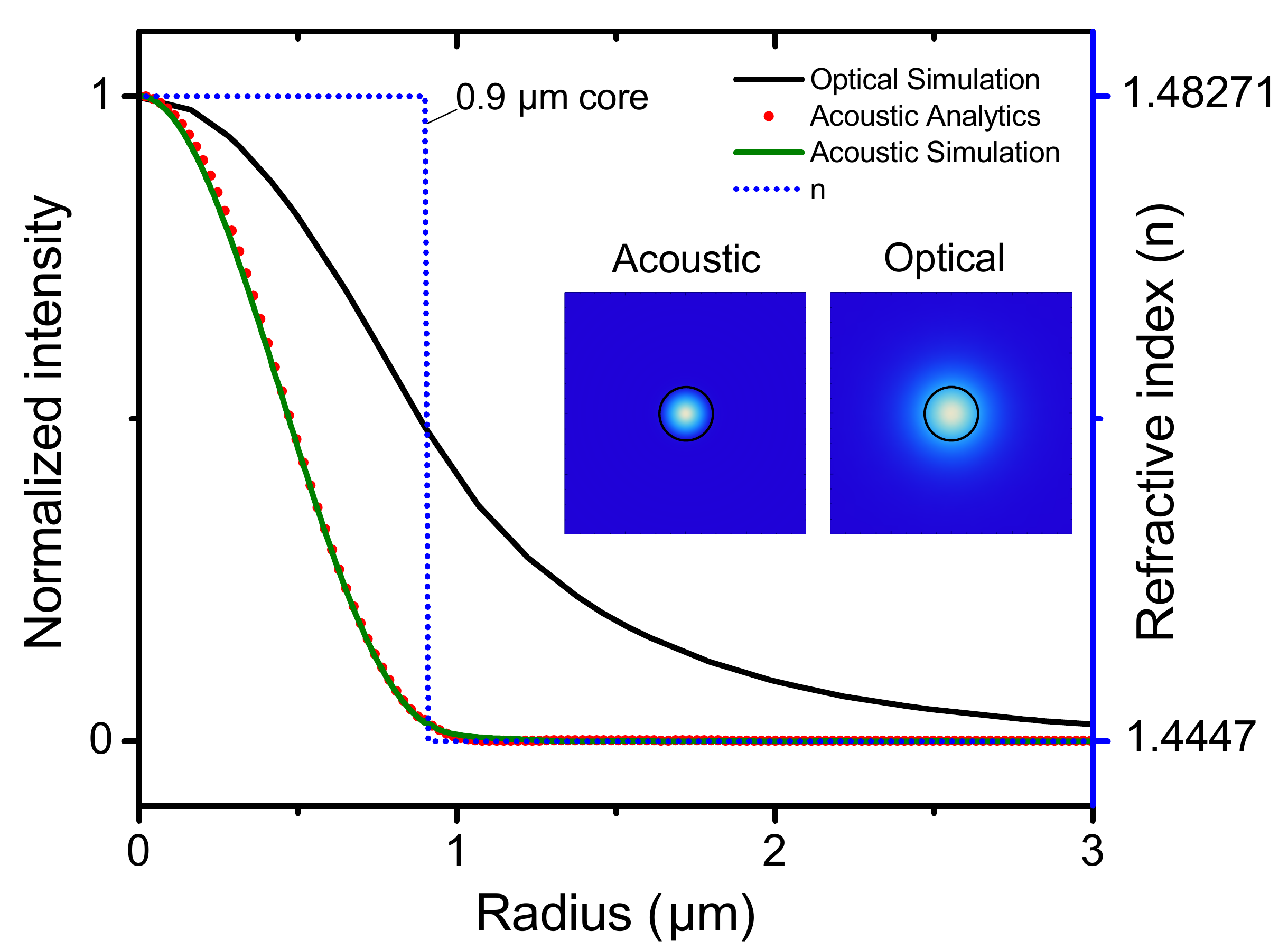}
\caption{Normalized fundamental mode acoustic and optical intensity, and index profile in UHNA3 fiber. 
Solid lines are obtained from finite element simulations, and red dots from analytical calculations of the acoustic modes.
Inset: Cross sectional view of the acoustic and optical intensity. The small core size and good opto-acoustic mode overlap ensures high optical and acoustic intensities as well strong optomechanical coupling.}
\label{SBS}
\end{center}
\end{figure}

Our system is a 2.2 cm segment of ge-doped Nufern UHNA-3 optical fiber. 
This fiber's high germanium concentration ($\sim 44 \ {\rm wt} \%$) guides light exceptionally
and produces a large longitudinal sound speed contrast between core $v_{\rm L,core} = 4,740 \pm 68$ m/s \cite{Dragic09} and cladding $v_{\rm L,clad} = 5,944$ m/s \cite{Jen86} 
that promotes acoustic guidance. The strong confinement and guidance of optical and acoustic modes (see Figs. \ref{TunnelingStates} \& \ref{SBS}) allows for the efficient excitation of phonons via stimulated 
Brillouin scattering (SBS) and which provides a means to characterize our system \cite{Ippen72,Boyd09,Agrawal95}.

Stimulated Brillouin scattering is a resonant process involving the interaction of light and sound. 
It occurs in media exhibiting photoelasticity and its complement electrostriction which are respectively 
characterized by a change in refractive index upon strain and by an optical intensity-induced elastic deformation.
Hence, the beat tone formed by two counter propagating optical fields can drive acoustic waves through 
electrostriction, and once created, acoustic waves act as moving Bragg gratings which reflect and Doppler shift 
incident optical beams as a consequence of photoelasticity. 
 
The concert of these two effects transfer energy between the two participating pump and Stokes optical fields, and the acoustic field with respective angular frequencies 
$\omega_\p$, $\omega_\s$, $\Omega$ and wavevector magnitudes $k_\p$, $k_\s$, $q$. 
The transfer of energy is only significant when the three fields are phase matched (see Fig. \ref{TunnelingStates} g-h); 
for backward SBS in 1D \cite{Boyd09} 
\ba
\omega_\p & = \omega_\s + \Omega \\
k_\p & = q - k_\s,
\end{align}
alternatively the phase-matching conditions can be viewed as energy and `momentum' conservation. 
The minus sign in front of $k_\s$ in the second equation above indicates that the Stokes photon and the pump 
counterpropagate (the phonon and the pump copropagate). For a given pump frequency we can approximate 
the phonon frequency, excited via SBS, by assuming linear dispersion for light ($\omega = (c/n) k$) and sound ($\Omega = v q$) 
\begin{equation}
\label{BrillouinFrequency}
\Omega \approx \frac{2 n v}{c} \omega_\p
\end{equation}
where $n$ is the effective index of refraction, $v$ is the modal sound speed, and $c$ is the speed of light. 

In our system SBS results in the transfer of energy from the pump to Stokes beam. 
Net Stokes amplification can be measured by balanced detection.  
For such measurements it is prudent to work in the weak signal regime, given by $g_\B P_\p L \ll 1$, where $g_\B$ is the Brillouin gain ($g_B \approx 0.6 $(Wm)$^{-1}$ at room temperature in UHNA-3 fiber), $P_\p$ is the pump power, and $L$ is the length of the 
fiber under test (FUT).
In this case the power transferred to the Stokes beam $\Delta P_\s$, which is directly acquired by balanced detection, is small and is given by
\begin{equation}
\label{PowerMeasured}
\Delta P_\s \approx \frac{ (\Gamma/2)^2  g_\B P_\p P_\s L }{ (\Omega -\omega_{\rm IM})^2+(\Gamma/2)^2}
\end{equation}
where we have dropped a negligible correction arising from the optical loss, $P_\s$ is the input power of the Stokes beam, the angular frequency $\omega_{\rm IM}$ is the angular frequency detuning between the pump and Stokes beams which is experimentally set by an intensity modulator (IM) see Fig. \ref{Apparatus},
and $1/\Gamma$ is the phonon lifetime. By sweeping $\omega_{\rm IM}$ balanced detection supplies the peak Brillouin gain, and the phonon frequency and lifetime which characterize SBS's Lorentzian frequency response.

\subsection{Phonon dissipation in glasses}

Dissipation of phonons in glasses has several origins: from multi-phonon interactions to scattering by defects. Among the latter of these are hypothesized TLSs that attenuate acoustic waves through the processes of resonant scattering and relaxation absorption. 
To understand the origins of these two mechanisms we lay out the tunneling state model and discuss its consequences for the dynamics of phonons.  For those seeking more details, complete derivations of all results in this section can be found in the Appendix. 

A tunneling state is characterized by a double-well potential of asymmetry $\Delta$ and the overlap energy $\Delta_0 = \hbar \omega_c e^{-\lambda}$ where $\hbar \omega_c$ is roughly the average zero point energy of the two wells and $\lambda = \frac{\sqrt{2 m V}d}{\hbar}$ characterizes the extent of wave function overlap between the two wells with $m$ the mass of the atom(s) comprising the TLS, $V$ the barrier height, $d$ the `distance' between the double-well's minima (see Fig. \ref{TunnelingStates}), and where $\hbar$ is Planck's constant divided by $2\pi$. Given a finite barrier height the atom(s) may tunnel between the two minima of the potential. The parameters $\Delta$ and $\lambda$ are assumed to be uniformly distributed over the ensemble of TLSs resulting in a constant density of states $G(\Delta,\lambda) = P$, or in terms of the overlap energy $f(\Delta,\Delta_0) = P/\Delta_0$.  This assumption of uniformity correctly predicts the linear in temperature behavior of the specific heat and the anomalous thermal conductivity of glasses at low temperatures \cite{Anderson72,Jackle72,Hunklinger82,Phillips87}.

At low temperatures the TLS+phonon system, including the strain-induced perturbation of the asymmetry, can be modeled with the following
Hamiltonian 
\begin{equation}
\label{Hamiltonian}
H = H_{\rm ph} + \sum_{j} \bigg[ \frac{1}{2}(E_j + D_{j} \cdot \xi({\bf r}_j)) \sigma_{z,j} + M_{j} \cdot \xi({\bf r}_j) \sigma_{x,j} \bigg]
\end{equation} 
\cite{Phillips87}
where the physics of the $j$th TLS is approximated as an effective two-level system using the Pauli matrices $\sigma_{i,j}$ with $i = \{x,y,z\}$.
The elastic strain field of polarization $\eta$ is labeled as $\xi_\eta$, the sum over $j$ counts all TLSs in the glass at various positions ${\bf r}_j$ with energy splitting $E_j$  
and coupling parameters $D_{\eta,j} \equiv 2(\Delta_j/E_j) \gamma_\eta$ and $M_{\eta,j}  \equiv (\Delta_{0,j}/E_j) \gamma_\eta$, $H_{\rm ph}$ is the free Hamiltonian for the phonons (see Appendix),
and the `dot product' $D_j \cdot \xi$ abstractly represents a sum over polarizations $ \sum_{\eta} D_{\eta,j} \xi_{\eta}$.
Generally, the deformation potential tensor, quantifying the TLS-strain coupling, is unique to each TLS and orientation dependent; we have ignored this complication above where the TLS-phonon interaction is characterized by a polarization dependent deformation potential constant $\gamma_\eta$. 
\subsection{Resonant phonon absorption by TLSs}
Tunneling states can resonantly interact with phonons though three processes: stimulated absorption, and spontaneous and stimulated emission. The relative magnitude of each of these processes determines the phonon dissipation rate and depends on temperature and the intensity of the acoustic field. 

\subsubsection{Weak fields}

First, we consider weak fields which is roughly characterized by a mean free time between TLS-phonon interactions that is long compared to the excited state lifetime of the TLS, and which will be defined quantitatively below. For such low intensity acoustic waves of angular frequency $\Omega$ the golden rule can be used to find the TLS-induced decay rate (see Appendix)
\begin{align}
\label{WeakFieldResonantAbs}
\Gamma^{\rm res} = & \frac{\pi P \gamma^2 \Omega}{ \rho v^2} \tanh \left(\frac{\hbar \Omega}{2 k_B T}\right) \\
                                           = & \frac{\pi P \gamma^2 \Omega}{ \rho v^2} (\underbrace{P_g}_{(i)} - \underbrace{P_e}_{(ii)}) \nonumber
\end{align}
\cite{Anderson72,Jackle72,Hunklinger82,Phillips87} where the polarization dependence of the deformation potential $\gamma$ and the sound speed $v$ have been suppressed, $k_B$ is the Boltzmann constant, and $\rho$ is the material density. 
$\Gamma^{\rm res}$ is characterized by two regimes occurring at high and low temperatures $T$. 
This is elucidated by the temperature dependence of the phonon decay rate on the thermal equilibrium population inversion $P_e - P_g$ of the TLSs 
at energy $\hbar \Omega$ and temperature $T$, where $P_e$ and $P_g$ are the probabilities
to find a TLS in the excited and ground state, respectively. At low temperatures $\hbar \Omega \gg k_B T$ the decay rate is maximized since the 
TLSs are found entirely in their ground state $(P_g \to 1)$, and thus $(i)$ stimulated absorption dominates. In the low intensity regime TLSs that have absorbed phonons decay through spontaneous emission which attenuates coherent phonon beams.  As the temperature is raised the probability to find the 
TLS in its excited state grows which opens the possibility for $(ii)$ stimulated phonon emission which coherently amplifies the sound amplitude. At high temperatures these two processes compensate each other exactly and resonant absorption is suppressed.  

\subsubsection{Strong fields}

At high acoustic intensities perturbation theory is no longer adequate to calculate the phonon decay rate. 
In this regime one must solve the coupled Heisenberg equations of motion for the TLSs and the acoustic field (see the Appendix for a complete derivation).
The backreaction of the TLSs on the phonons is characterized by a complex susceptibility
that modifies the phonon dynamics in two ways: by resonant absorption induced dissipation and a  
frequency shift. At high acoustic intensities the dissipation rate due to resonant absorption is given by 
\begin{equation}
\label{ResAbsorption}
\Gamma^{\rm res} \approx   \frac{\pi P \gamma^2 \Omega}{\rho v^2} \frac{\tanh(\hbar \Omega/2 k_B T)}{\sqrt{1+J/J_c}}.
\end{equation}
\cite{Phillips87} where $J$ is the acoustic intensity, and $J_c$ is the critical intensity that demarcates the boundary between weak and strong fields. 
The critical intensity 
\begin{equation}
\label{criticalIntensity}
J_c  \approx \frac{\hbar^2 \rho v^3}{2 \gamma^2 T_1 T_2}
\end{equation}
is characterized by the time scales $T_1$ and $T_2$ which 
are phenomenological upper state lifetime and dephasing times for the TLSs with energy $E \approx \hbar \Omega$ (see Appendix).

As with the case of weak fields, this result for the phonon dissipation can be interpreted as a competition between stimulated emission and absorption. 
In the strong field case however the effective population inversion for the TLSs which interact with a phonon of frequency $\Omega$ is given by the 
nonequilibrium steady-state value 
\begin{equation}
\label{NEqPopInv}
P_e - P_g = -\frac{\tanh(\hbar \Omega/2 k_B T)}{\sqrt{1+J/J_c}}. 
\end{equation}
At high intensities $J/J_c \gg 1$ a TLS, excited through resonant absorption, 
will encounter several phonons within its upper state lifetime. Thus, stimulated emission 
can become a dominant decay channel and resonant absorption will be suppressed. 

Resonant absorption also leads to a (acoustic intensity-insensitive) frequency shift given by
\begin{align}
\label{frequencyShift}
\Delta \omega^{\rm res}(T) - & \Delta \omega^{\rm res}(T_0) \approx   - \frac{P \gamma^2 \Omega}{\rho v^2}
\bigg[\ln \left(\frac{\hbar \Omega}{k_{ B} T} \right) \nonumber 
\\
& -  {\rm Re} \Psi \left( \frac{1}{2} + \frac{\hbar \Omega}{2\pi i k_{B} T} \right) \bigg] -(T \to T_0),
\end{align}
where $\Psi$ is the digamma function \cite{Phillips87}. 

\subsection{Relaxation absorption}

In addition to resonant absorption TLSs can attenuate coherent acoustic waves via relaxation absorption. 
This process results from a modulation of the TLS's energy splitting by the presence of a time-dependent strain field, i.e. $E_j(t) \to E_j + D_{\eta,j}\xi_\eta(t)$. 
As $E_j(t)$ changes, the TLS can equilibrate with the surroundings by absorbing or releasing energy. Thus, the instantaneous `equilibrium'  population inversion
is modulated in time. For our system (see Appendix) relaxation-absorption results in dissipation of phonons characterized by the decay rate
\begin{equation}
\label{RelAbsorption}
\Gamma^{\rm rel} \approx \frac{\pi^3}{24} \frac{P \gamma_\eta^2}{\rho^2 v_\eta^2 \hbar^4} \left( \sum_{\eta'} \frac{\gamma^2_{\eta'}}{v_{\eta'}^5}\right) (k_B T)^3,
\end{equation}
and a temperature-independent frequency shift.  


\section{Experimental Setup} 
\begin{figure}[t]
\begin{center}
\includegraphics[width=3.3in]{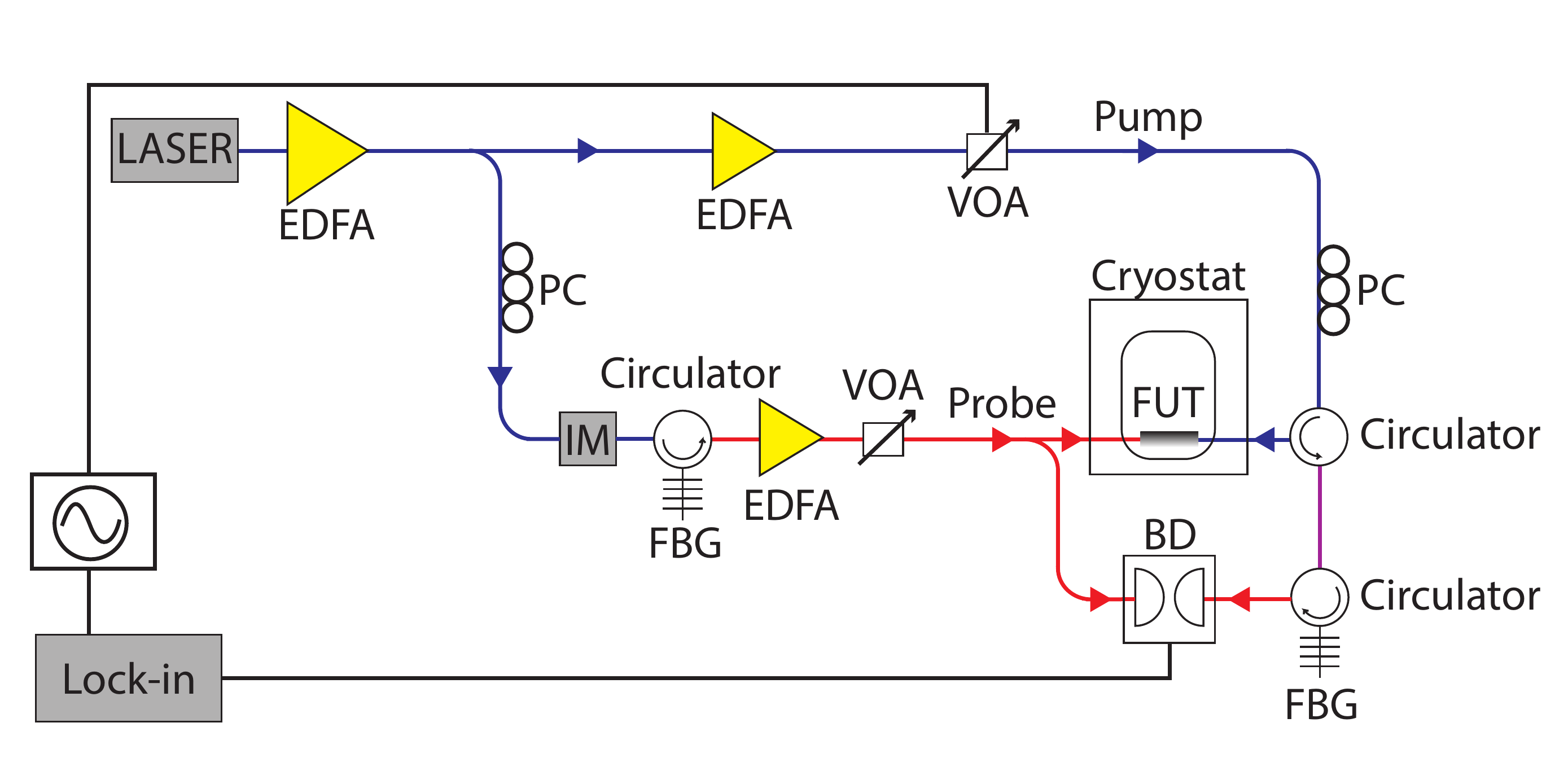}
\caption{Experimental setup to measure SBS with balanced detection. The acronyms above are defined as: EDFA Erbium doped fiber amplifier, VOA variable optical attenuator, RF radio frequency signal generator, BD balance detector, and FBG fiber Bragg grating. We use a DFB (distributed feedback) laser with a wavelength of 1548.963 nm and a linewidth of 5 kHz. Our probe line is synthesized by sending pump laser light through an IM modulated in the neighborhood of 9.2 GHz, the carrier and the high frequency side-band are removed by operating at the null point bias point of the IM and  band-pass filtering. Probe light is split in two arms, one acting as a reference and the other traversing the FUT. The power modulated pump results in a time harmonic amplification of the Stokes beam at $\Omega_{\rm mod}$ which is acquired with lock-in detection.}
\label{Apparatus}
\end{center}
\end{figure}

The system is characterized by pump-probe measurements performed using balanced detection and lock-in amplification. Our apparatus consists of a $1548.963$ nm source that is split into two optical lines (see Fig. \ref{Apparatus}); one reserved to act as a pump, and the second to act as a probe. The pump line is subsequently amplified and pump and probe polarizations are aligned. The pump beam is then power modulated at the fixed frequency $\Omega_{\rm mod}$ for lock-in detection before being sent through the FUT. The probe beam is sent through an intensity modulator which upon exit is filtered to isolate a single side-band. To achieve common mode noise rejection using balanced detection, the probe beam is split into two arms. One arm passes probe light through the FUT and is amplified via SBS whereas the other arm acts as a reference. A variable attenuator on the reference arm is adjusted so that both probe arms have the same optical power in the absence of gain via SBS.  The difference in power of the balanced probe arms yields the net gain experienced by the probe beam passing through the sample.

Modulation of the pump power generates a fixed-frequency side-band of the amplified probe which is detected with a lock-in amplifier. To measure Brillouin gain spectra (BGS) the modulation frequency $\omega_{\rm IM}$ is swept across the Brillouin resonance generating a Lorentzian response. 

The temperature of the fiber is controlled using a double walled cryostat, and a large copper heat sink. The samples are mounted in a shallow mail slot passing through the block. The block size was chosen to ensure temperature stability, and to allow a minimum of 10 Brillouin lineshape measurements per 100 mK as the block warms slowly to room temperature. The system is cooled using liquid Helium and evaporative cooling resulting in lowest achievable temperatures in the neighborhood of 1 K. 

To study the saturation of losses arising from the TLSs the intensity of the sound field inside the FUT was swept at fixed temperature. In the weak signal regime the steady-state phonon intensity follows the powers of the optical fields
\begin{equation}
\label{PhononIntensity}
 J(\omega_{\rm IM}) \approx \frac{v}{\Gamma} \frac{\omega_{\rm IM}}{\omega_\s} \frac{1}{A_{\rm eff}} G_{\rm B}(\omega_{\rm IM}) P_\p P_\s ,
\end{equation} 
where $A_{\rm eff} \approx 1.6 \mu$m$^2$ (estimated from simulations, see Fig. \ref{SBS}) is the phonon mode area. Hence, the phonon intensity can be controlled by changing the power of the optical driving fields. By sequential adjustment of the variable optical attenuators on pump and probe lines the phonon intensity generated through SBS can be swept through more than 4 decades of dynamics range. The maximum pump and probe powers were used to determine the 2.2 cm length of the FUT;
this length ensures that all measurements were performed in the weak-signal regime. This fiber length allowed access to high phonon intensities while simultaneously preventing inhomogeneous broadening from non-linearity induced by strong backward scattering via SBS.


\section{Results}

\subsection{Temperature and Intensity Dependence of Phonon Losses}

\begin{figure*}[ht]
  \centering \includegraphics[width=7in]{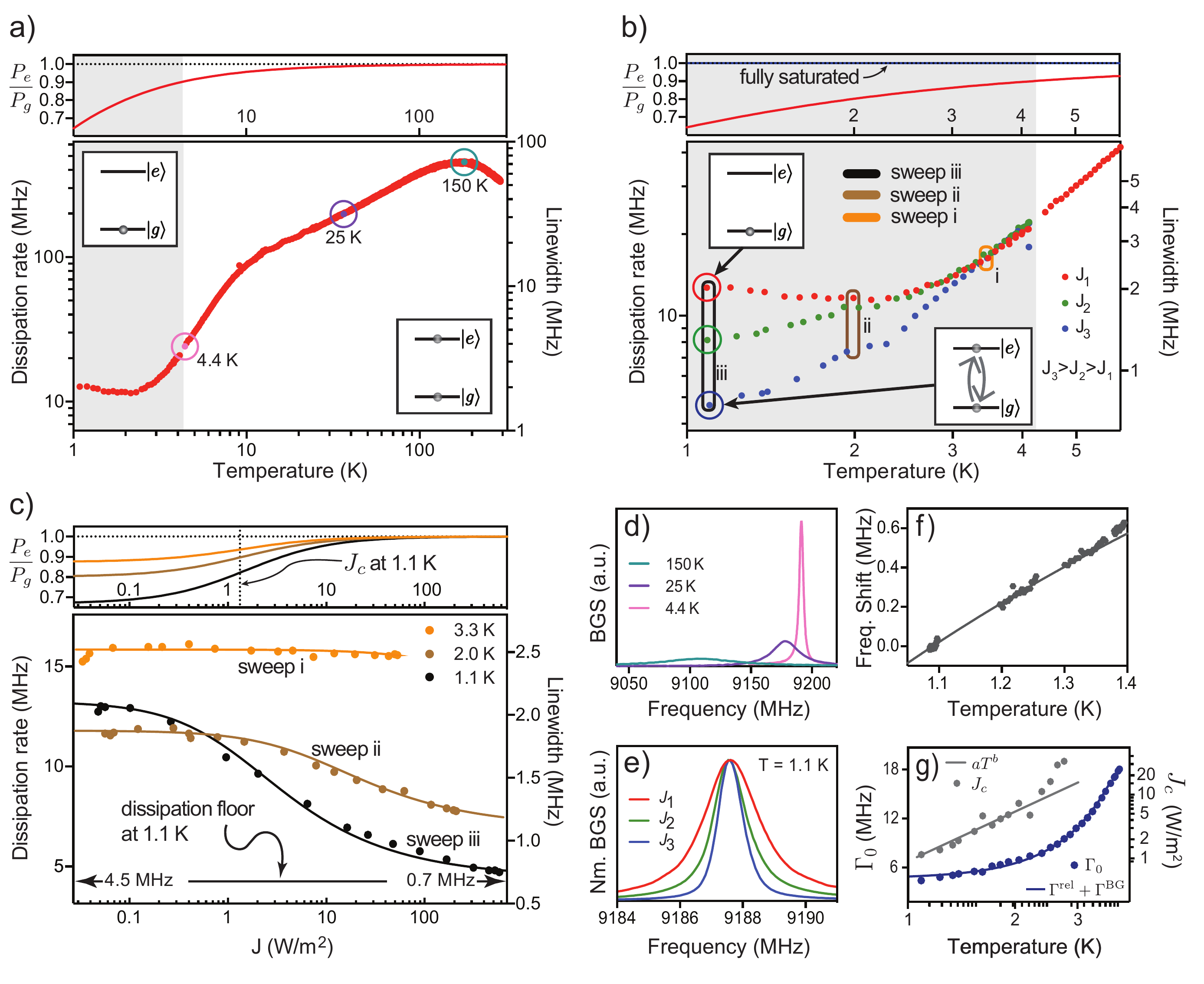}
  \caption{
  a) Dissipation rate of the fundamental acoustic mode at low intensity as a function of temperature.  
  b) Phonon dissipation rate as a function of temperatures below 6 K. The three data sets below 4 K correspond with different pump and probe power settings corresponding with low, moderate, and high phonon intensities, as compared to $J_c$.
  c) Phonon dissipation rate as a function of acoustic intensity. Each curve has a fixed temperature.
  d) Brillouin gain spectra as a function of temperature. 
  e) Normalized Brillouin gain spectra as a function of intensity at 1.1 K.
  f) Comparison of the measured Brillouin frequency vs. temperature with the theoretical prediction given by Eq. \ref{frequencyShift}. 
  g) Critical intensity $J_c$ and background losses $\Gamma_0$ as a function of temperature. 
 The parameters for the theory can be found in Tab.1}
  \label{Results-I}
\end{figure*}

Brillouin gain spectra corresponding to the fundamental acoustic mode in a 2.2 cm segment of UHNA-3 fiber were continuously acquired as the system slowly warmed to room temperature from 4.17 K. Pump and probe powers were set to 35 mW and 0.550 mW, respectively to ensure SBS measurements in the weak signal regime. The BGS were binned in 100 mK steps, averaged, and fit to the Lorentzian model given by Eq. \ref{PowerMeasured}. This analysis provides the phonon frequency $\Omega$ and the dissipation rate $\Gamma$ as a function of temperature. 
The red data points in Figure \ref{Results-I}a show the dissipation rate of the fundamental acoustic mode as a function of temperature. Three representative BGS are shown in Fig. \ref{Results-I}d. To judge the relative importance of resonant absorption by TLSs we have plotted the ratio $P_e/P_g$ as a red curve (top).
 
For low phonon intensities ($J \ll J_c$) the linewidth begins to level off below 4K (red points in Figs. \ref{Results-I}a \& \ref{Results-I}b) and then begins to increase as the temperature is lowered. This gray region in Figs. \ref{Results-I}a \& \ref{Results-I}b, where $P_e/P_g < 0.9$, indicates the temperature range where resonant absorption by TLSs begins to dominate the acoustic damping.   

After cooling the system to 1.1 K, we let the fiber system slowly warm up to 4.17 K. For each 100 mK rise in temperature we acquire BGS as a function of phonon intensity. The data is analyzed by binning and averaging as described above, and results in dissipation rate as a function of both temperature and phonon intensity. This data is presented in \ref{Results-I}b \& \ref{Results-I}c: \ref{Results-I}b shows phonon dissipation rate as a function of temperature. The three sets of data correspond with three distinct settings of the optical powers that generate phonons of low, moderate, and high intensities (as compared to $J_c$). Normalized BGS for the three settings is shown in Fig. \ref{Results-I}e for the lowest temperature of 1.1 K. 
Figure \ref{Results-I}c shows phonon dissipation rate as a function of phonon intensity for three different temperatures (black, orange and tan points). It is qualitatively clear from Figs. \ref{Results-I}b \& \ref{Results-I}c that at temperatures lower than 4 K phonon dissipation is suppressed at high phonon intensity; this indicates saturation of the resonant absorption due to TLSs and is made evident by the theoretical plot of $P_e/P_g$ (at top), computed using Eq. \ref{NEqPopInv}.

In Figs. \ref{Results-I}c, f, \& g we compare our measurements with the tunneling state model. In Fig. \ref{Results-I}c a model of the phonon dissipation rate given by 
\begin{equation}
\label{ }
\Gamma = \Gamma^{\rm res} + \Gamma_0
\end{equation}
is fit to the data where $\Gamma_0$ is an offset parameter that represents all intensity independent background acoustic losses including relaxation absorption, Rayleigh scattering, phonon-phonon interactions, etc. Three parameters are employed to fit the data: 
1)  $P \gamma^2_L$ ($\gamma_L$ being the deformation potential for longitudinal waves), 
2)  $J_c$, and 3) the offset $\Gamma_0$. The parameters $P$ and $\gamma_L$ are assumed to be constant. 
In distinction, $J_c$ and $\Gamma_0$ are expected to depend upon temperature. In particular, $J_c$ scales inversely with the effective decay rates $T_1 T_2$.  The upper state lifetime $T_1$ can be approximated with the Fermi's golden rule (see Eq. \ref{TLSlifetime} of the Appendix), but an estimation of the temperature dependence of $T_2$ is beyond the scope of this work. Instead we fit $J_c$ (gray dots in Fig \ref{Results-I}g) to a power law $J_c = a T^b$ that is shown as a solid gray line in \ref{Results-I}g. 
In Fig. \ref{Results-I}f the measured shift in Brillouin frequency (gray points) referenced to $T_0 = 1.09$ K is compared to Eq. \ref{frequencyShift}. The fitted values obtained from Fig. \ref{Results-I}c were used as inputs. 
The offset $\Gamma_0$ is plotted as blue points in Figs. \ref{Results-I}g and contains all background losses which we model as 
\begin{equation}
\label{ }
\Gamma_0 = \Gamma^{\rm rel} + \Gamma^{\rm BG}. 
\end{equation}
where $\Gamma^{\rm BG}$ is a constant representing the background dissipation floor for our system. The model is fit to the data by adjusting the offset and the value of the longitudinal deformation potential $\gamma_L$ (we assume that the transverse deformation potential is given by $\gamma_T^2 \approx  \gamma^2_L/2$ \cite{Golding76b}). $P \gamma^2_L$ is held fixed to the value obtained from the data analysis in Fig. \ref{Results-I}c. 

The model parameter values obtained from this analysis are tabulated in Tab. \ref{FittedParameters} and their values are compared with those of vitreous silica. There are several remarks in order. The data of Fig. \ref{Results-I}c is well-described by the tunneling state model, and the computed  Brillouin frequency shift using the fitted value of $P \gamma_L^2$ from Fig. \ref{Results-I}c compares well with our measurements, however the theory and experiment begin to diverge at higher temperatures. Relaxation absorption accounts well for the intensity independent background, and the fitted value for $\gamma_L$ is comparable to silica. The fitted  parameters and their theoretical relationships can also be used to estimate the density of TLSs $P$ and the dephasing rate $T_2$. 

As a final note, the values of $T_1$ and $T_2$ should not be directly compared with those of vitreous silica listed in Tab. \ref{FittedParameters}. The phonon frequencies and the temperatures were distinct from our measurements; for an appropriate comparison $T_1$ can be estimated from the minimum TLS upper state lifetime computed using perturbation theory Eq. \ref{TLSlifetime} with $\Delta_0 \to E$, taking $T = 20$ mK and $\Omega/2\pi$ to be $0.68$ GHz, and using the fitted TLS parameters for ge-doped silica. We find $T_1 \approx 665$ $\mu$s which is comparable to the reported values for vitreous silica in Tab. \ref{FittedParameters}

\begin{table}
  \centering 
\begin{center}
    \begin{tabular}{ | l | l | l | p{2.0in} |}
    \hline
				& $44 \%$ wt. ge-doped silica 				& vitreous silica  						\\ \hline
    $\rho$ 			& $2,666$ kg m$^{-3}$ 			& $2,202$ kg m$^{-3}$	 \cite{Jen86}					\\ \hline
    $v_L$ 			& $4,760 \pm 68$ m s$^{-1}$ 	\cite{Dragic09}				& $5,944$ m s$^{-1}$  	 \cite{Jen86}					\\ \hline
    $a$ 			& $0.9$ W m$^{-2}$ K$^{-b}$  		&  											\\ \hline
    $b$ 			& $2.6$ 						&  											\\ \hline
    $J_c$ 			& $1.2$ W m$^{-2}$ 			&  											\\ \hline
    $P \gamma_L^2$ 	& $1.6 \times 10^7$ J m$^{-3}$ 		&   $1.3 \times 10^7$ J m$^{-3}$	\cite{Graebner86}		\\ \hline
    $P$ 			& $23$  	&  $6.85$ 		\cite{Graebner86}									\\ \hline
    $\gamma_L$ 		& $0.5$ eV  					&  $0.86$ eV	\cite{Graebner86}								\\ \hline 
     $\sqrt{T_1 T_2}$ 	& $ 10$ ns (est. with Eq.\ref{criticalIntensity})				&  $^*$$53$ $\mu$s  \cite{Golding76}				\\  \hline
    $T_1$ 			& $\approx 79$ ns (est. with Eq.\ref{TLSlifetime})				&  $^*$$200$ $\mu$s  \cite{Golding76}			\\  \hline
    $T_2$ 			& $\approx 1.3$ ns 																&  $^*$$14$ $\mu$s	 \cite{Golding76}									\\  \hline
    \end{tabular}
\end{center}
\caption{Comparison of the properties of ge-doped silica and silica. All temperature and frequency dependent quantities are listed for $T = 1.1$ K and $\Omega = (2\pi) 9.188$ GHz. The product $T_1 T_2$ is extracted from $J_c$ defined by Eq. \ref{criticalIntensity}.
The parameter $T_1$ above is calculated from  Eq. \ref{TLSlifetime}, and then subsequently employed to extract $T_2$ from $J_c$. 
$P$ is in units of $10^{44}$ J$^{-1}$ m$^{-3}$.
$^*$The values of $T_1$ and $T_2$ for vitreous silica were measured at $T = 20$ mK and for frequency $0.68$ GHz. 
Hence, these values should not be directly compared the values listed for ge-doped silica. Using Eq. \ref{TLSlifetime} and the fitted tunneling state parameters
$T_1$ at 20 mK and for 0.68 GHz phonons is estimated at 665 $\mu$s in UHNA-3, this compares well with the listed value for vitreous silica. 
}
\label{FittedParameters}
\end{table}


\section{Discussion} 

Acoustic dissipation imposed by defects presents a barrier to attaining low loss acoustic modes that are critical to a range of technologies. To overcome/understand these challenges we explored defect-induced dissipation in a guided wave system using nonlinear phonon spectroscopy. Theory-experiment comparison using the tunneling state model quantifies the influence of defects in our system. We found that the large phonon intensities, made possible by the tight acoustic confinement, permit access to the TLS-induced regime of nonlinear phonon dynamics. We demonstrated self-frequency saturation, where TLS resonant absorption is driven to transparency as the intensity of the acoustic beam becomes large.

At the highest intensities we encounter a dissipation floor $\Gamma_0$ due to relaxation absorption $\Gamma^{\rm rel}$ and a roughly temperature independent offset $\Gamma_{\rm BG}$. At the lowest temperatures and highest intensities we estimate that relaxation absorption makes up only 6\% of the dissipation rate floor, and thus we believe we have reached the neighborhood of the smallest phonon dissipation rates possible in UHNA3 fiber estimated to be $(2\pi)650$ kHz.There are several possible explanations for $\Gamma_{\rm BG}$ such as; phonon-phonon scattering, anchor losses, inhomogeneous broadening due to irregularities in the fiber geometry along its length, acoustic leakage due to imperfect guiding, and Rayleigh scattering due to disorder and scattering by dopants.  Phonon-phonon scattering can be estimated from the Landau-Rumer theory using Fig. 2 of \cite{Goryachev13c} and gives negligible dissipation, and anchor losses are likely unimportant as the mode is highly confined to the core. Large amounts of inhomogeneous broadening are unlikely given the relatively short length of the fiber and the observed Lorentzian shape of the BGS, and acoustic leakage was estimated in simulations to be too small. 

We attribute the dissipation floor to Rayleigh scattering. There are several pieces of corroborating evidence for this hypothesis;
the fiber is doped with a large concentration of germanium that will produce large stochastic variations of density and sound speed in the fiber core,
$\Gamma_{\rm BG}$ is insensitive to temperature and intensity, and an estimation of Rayleigh scattering in Top High Quality alpha synthetic quartz  
for $9.2$ GHz phonons gives a lower bound for the Rayleigh scattering dissipation rate $(2\pi)200$ kHz.  By working at lower frequencies (as Rayleigh scattering scales with the fourth power of the frequency) or with fibers with lower defect densities much longer phonon lifetimes may be possible. 


Acknowledgements: Primary support for this work provided by NSF grant DMR 1119826. 

\appendix

\section{Acoustic guidance in fiber}

The high germanium concentration ($\sim 44 \ {\rm wt} \%$) in UHNA-3 fiber produces a large longitudinal sound speed contrast between core $v_{\rm L,core} = 4,740 \pm 68$ m/s \cite{Dragic09} and cladding $v_{\rm L,clad} = 5,944$ m/s \cite{Jen86}
that promotes acoustic guidance. However, ideal guidance is not expected since the shear wave velocity in the cladding $v_{\rm S,clad} = 3,764$ m/s \cite{Jen86} is smaller than $v_{\rm L,core}$, and hence leads to mode conversion at the core-clad interface \cite{Rose99} that leads to acoustic leakage. 

To understand the limits of acoustic guidance we studied the axial-radial acoustic modes of our system with full vectorial simulations and semi-analytical calculations. We solved for acoustic eigenfrequencies of a core region (diameter 1.8 $\mu$m) embedded in a finite cladding (diameter 125 $\mu$m) surrounded by liquid helium. A perfectly absorbing boundary condition was implemented adiabatically in the liquid region at large separation from the core to prevent spurious reflections from the simulation boundaries. 
We employed a shear velocity in the core region $v_{\rm S,core} \sim 3092$ m/s which was estimated by interpolating between the values of pure silica and germania, and the liquid region was ascribed the acoustic properties of liquid helium. The core, cladding and liquid were assumed lossless other than the absorbing layer used to model mode leakage. 

Acoustic energy that enters the liquid region leaves the system irreversibly and hence reduces the acoustic guidance. The signature of this leakage is an imaginary part to the eigenfrequencies that quantifies the dissipation rate. Our simulations correctly predict the acoustic frequency and yield negligible leakage (much smaller than the dissipation rate observed in experiment). Hence, we conclude that the dissipation of phonons in our system is due almost entirely to internal friction of the fiber as opposed to leakage. In this sense we describe our phonons as well-guided. 

It should be emphasized that acoustic guiding in fiber is not guaranteed. Germanium doping has the fortunate consequence that increased doping leads to higher index of refraction while simultaneously decreasing sound speed \cite{Jen86,Kong12}, thus enhancing the confinement of light and sound simultaneously. Titanium dioxide and phosphorous pentoxide are other dopants which simultaneously enhance guiding of light and sound, while boron trioxide and fluorine doping enhance the confinement of light while raising the sound speed \cite{Thurston92}. Fibers doped with the former will possess leaky acoustic modes. At room temperature, where phonons are strongly damped in silica, one may posit that the guiding of sound in fiber plays a minor role as the mean free path is on the order of the acoustic mode field diameter. However, the existence of higher order guided phonon modes can be observed at room temperature \cite{Shelby85,Kobyakov09} which shows that acoustic modes in fiber cannot be completely understood in terms of bulk properties without considering boundaries. We have also simulated axial-radial modes in a system composed of a core embedded in an infinite cladding region. In this case the computed acoustic leakage rate exceeds the observed phonon dissipation rate suggesting that the cladding liquid interface plays an important role in the observed acoustic guidance. In light of this fact it is important note that the leakage rate we derived from simulation for the finite fiber case is only a lower bound as we have not accounted for nonuniformity in the geometry of the fiber over its length. Irregularities in cladding and core diameters will contribute to inhomogenous broadening that will manifest as an effective increase in the leakage rate. 

\section{Tunneling state theory of phonon dissipation in glasses}

In this section we summarize the underpinnings of the tunneling state theory.

At low temperatures $k_B T \ll \hbar \omega_c$ a TLS can be effectively represented by the Hamiltonian 
\begin{equation}
\label{TLSHamiltonian}
H_{\rm TLS} = \frac{1}{2} \begin{pmatrix}
\Delta      &   \Delta_0 \\
\Delta_0      &   - \Delta
\end{pmatrix}
\end{equation}
where the states $\left< L \right| = (0,1)$ and $\left< R \right| = (1,0)$ correspond with position states with the particle localized in the left (`$L$') or right (`$R$') well and which we approximate as being orthogonal. 
Diagonalizing the Hamiltonian above gives the corresponding energy eigenstates  
\begin{align}
\label{}
\left| e \right>    &  = \frac{\Delta_0}{\sqrt{2E (E +  \Delta)}}\bigg[ \left| L \right> + \frac{\Delta + E}{\Delta_0} \left| R \right> \bigg] \nonumber \\
\left| g \right>    &  = \frac{\Delta_0}{\sqrt{2E (E -  \Delta)}}\bigg[ \left| L \right> + \frac{\Delta - E}{\Delta_0} \left| R \right> \bigg] 
\end{align}
which reveal that the stationary states of a free TLS are spatial superpositions between the wells and have energies $\pm \frac{1}{2}E \equiv \pm \frac{1}{2}\sqrt{\Delta^2 + \Delta_0^2}$, see Fig. \ref{TunnelingStates}b.
Such a TLS is perturbed by a local strain; the dominant effect of this is a shift in the asymmetry $\Delta$ that perturbs the diagonal elements of the Hamiltonian in Eq. \ref{TLSHamiltonian} as $\Delta \to \Delta + \frac{\partial \Delta}{\partial \xi_{\rm ab}} \xi_{\rm ab}$ where $\xi_{\rm ab}$ is the elastic strain, $\frac{1}{2} \frac{\partial \Delta}{\partial \xi_{\rm ab}} \equiv \gamma_{\rm ab}$ is the deformation potential characterizing the linear response of the TLS potential to strain, `a' and `b' denote spatial components of the strain tensor, and the Einstein summation convention for repeated indices is used. Upon applying the unitary transformation which diagonalizes the Eq. \ref{TLSHamiltonian} to the full Hamiltonian including the strain-induced perturbation of the asymmetry and averaging over TLS orientations, we arrive at Eq. \ref{Hamiltonian}. The free Hamiltonian for the phonons is given by 
\begin{equation}
\label{ }
\sum_{{\bf q},\eta} \hbar \Omega_{{\bf q}\eta}(b^\dag_{{\bf q}\eta} b_{{\bf q}\eta} +1/2)
\end{equation}
where the sum counts modes with wavevector and polarization ${\bf q}$ and $\eta$, and $b_{{\bf q}\eta}$ and $\Omega_{{\bf q}\eta}$ are the ${\bf q}\eta$-mode annihilation operator and frequency, respectively.  
The decomposition of the strain field into normal modes, here in plane waves, reveal the connection with the annihilation operator $b_{{\bf q}\eta}$
\begin{align}
\label{strain}
\xi_{\rm ab}({\bf x}) = &\frac{i}{2} \sum_{{\bf q},\eta} (q_{\rm a} \hat{r}_{\eta b}({\bf q})  + q_{\rm b} \hat{r}_{\eta a}({\bf q}) )
\sqrt{\frac{\hbar}{2 \Omega_{{\bf q}\eta} \rho V}}  e^{i {\bf q} \cdot {\bf x}} b_{{\bf q}\eta} \nonumber \\
& +H.c.
\end{align}
where $V$ is volume of the system, $\rho$ is the material density, $\hat{r}_{\eta b}({\bf q})$ is a unit vector for $\eta$-polarized phonons, and $H.c.$ stands for Hermitian conjugate. 

\section{TLS excited state lifetime}

Before we begin our investigation of TLS-induced effects upon the phonons 
we outline some of the phonon-induced effects upon the TLSs which play a
role in our  analysis. The most important of these effects is that interaction 
with the phonons leads to a finite TLS upper state lifetime. 
 
For a TLS of asymmetry $\Delta$ and overlap energy $\Delta_0$ Fermi's golden rule gives the excited state decay rate
\begin{equation}
\label{TLSlifetime}
\frac{1}{\tau} = \sum_\eta \frac{\gamma^2_\eta}{v_\eta^5}\frac{ E \Delta_0^2}{2\pi \rho \hbar^4} \coth \left(\frac{E}{2 k_B T}\right) 
\end{equation}
\cite{Anderson72,Jackle72,Hunklinger82,Phillips87} 
where the sum over $\eta$ counts decay channels corresponding with the various phonon polarizations, and $T$ is the temperature of the phonon bath. The polarization dependence of the deformation potential and the sound speed are accounted for in the $\eta$ suffices. 
For fixed energy $E$ the lifetime $\tau$ has the minimum value $\tau_{\rm min}$ obtained from Eq. \ref{TLSlifetime} by taking $\Delta_0 \to E$. 

The lifetime of the $j$th TLS is derived using Fermi's golden rule. 
We begin with matrix elements for upward transition of the TLS from the ground $\left| g_j \right>$ to the excited state
$\left| e_j \right>$ 
\begin{align}
\label{ }
\left< e_j,n_{\bf q\eta}-1\right| & M_j:\xi({\bf r}_j) \sigma_{x,j}  \left| g_j, n_{\bf q\eta} \right> 
= \nonumber \\
& i q \gamma_\eta \sqrt{\frac{\hbar}{2 \Omega_{\bf q\eta} \rho V}} e^{i {\bf q}\cdot {\bf r}_j} \frac{\Delta_0}{E} \sqrt{n_{\bf q\eta}}
\end{align}
where $n_{\bf q\eta}$ is the number of phonon quanta in the ${\bf q}\eta$-mode. 
The matrix element for downward transitions is given by
\begin{align}
\label{ }
\left< g_j,n_{\bf q\eta}+1\right| & M_j:\xi({\bf r}_j) \sigma_{x,j}  \left| e_j, n_{\bf q\eta} \right> 
= \nonumber \\
 &- i q \gamma_\eta \sqrt{\frac{\hbar}{2 \Omega_{\bf q\eta} \rho V}} e^{-i {\bf q}\cdot {\bf r}_j} \frac{\Delta_0}{E} \sqrt{n_{\bf q\eta}+1}.
\end{align}
After averaging over the initial (thermal) phonon state and summing over all final states that contribute to the two processes
the golden rule gives the upper transition rate as
\begin{equation}
\label{ }
C_{g\to e} = \sum_\eta \frac{\gamma_\eta^2}{v^5_\eta} \frac{ E_j \Delta_{0,j}^2}{2\pi \rho \hbar^4} \frac{1}{e^{\frac{E_j}{k_B T}}-1}
\end{equation}
where the Debye phonon density of states has been used $g_\eta(E) = E^2/(2\pi^2 \hbar^3 v^3_\eta) V$. The rate $C_{e \to g}$ can 
similarly be computed and is identical to $C_{g \to e}$ if the factor $(\exp (E_j/k_B T)-1)^{-1}$ is taken to $(\exp(E_j/k_B T)-1)^{-1}+1$. 

These two rates can be combined to give the time rate of change for probability of the TLS to be in its excited state
\begin{align}
\label{ }
\dot{P_e} = & C_{g \to e} P_g - C_{e \to g} P_e \nonumber \\
                = & -(C_{g \to e} + C_{e \to g}) P_e + C_{g \to e}
\end{align}
since $P_e + P_g = 1$ which gives the decay rate for the excited state as $C_{g \to e} + C_{e \to g}$ written explicitly in Eq. \ref{TLSlifetime}. 

\section{Resonant phonon absorption by TLSs}

\subsection{Derivation of weak field phonon decay rate}

Eq. \ref{WeakFieldResonantAbs} is derived using the golden rule. The transition rates for an increase in the number of quanta in the ${\bf q \eta}$-phonon mode 
from $n_{\bf q \eta}$ to $n_{\bf q \eta}+1$ as well as the decay rate from $n_{\bf q \eta}$ to $n_{\bf q \eta}-1$ are given by
\begin{align}
\label{ }
C_{n_{\bf q \eta} \to n_{\bf q \eta}+1} =  \frac{1}{V} \sum_j P_e \frac{\Delta_{0,j}^2 \gamma_\eta^2}{E^2_j} \frac{\pi q^2}{ \Omega_{\bf q \eta} \rho}  (n_{\bf q \eta}+1) \delta(E_j-\hbar \Omega_{\bf q \eta}) \nonumber \\
C_{n_{\bf q \eta} \to n_{\bf q \eta}-1} = \frac{1}{V} \sum_j P_g \frac{\Delta_{0,j}^2 \gamma_\eta^2}{E^2_j} \frac{\pi q^2}{ \Omega_{\bf q \eta} \rho}  n_{\bf q \eta} \delta(E_j-\hbar \Omega_{\bf q \eta}).
\end{align}

The sum over the various TLSs can be performed by assuming the validity of the ergodic theorem 
which states that the volume average is equal to the ensemble average of the TLSs in the thermodynamic limit
i.e. $\frac{1}{V} \sum_j \mathcal{F}(\Delta_j,\Delta_{0,j},E_j)\to \int d\Delta \int d \Delta_0 P/\Delta_0 \ \mathcal{F}(\Delta,\Delta_{0},E)$ 
where $\mathcal{F}$ represents the summand and recall that $P/\Delta_0$ is the TLS density of states. 

Converting the sum over $j$ to an integral over the TLS distribution and completing the integrals over $\Delta$ and $\Delta_0$
gives time rate of change of $n_{\bf q \eta}$ 
\begin{align}
\label{ }
\dot{n}_{\bf q \eta}  = & \lambda  [P_e(\underbrace{n_{\bf q \eta}}_{\rm (i)}+\underbrace{1}_{\rm (ii)})-P_g \underbrace{n_{\bf q \eta}}_{\rm (iii)}]
\end{align}
where $\lambda(P_e - P_g)$ is given by Eq. \ref{WeakFieldResonantAbs}.  
The underbraces denote the terms contributing to; (i) stimulated phonon emission, (ii) spontaneous phonon emission, and 
(iii) stimulated phonon absorption. From the expression above it is clear that when $P_g \to 1$ that the TLSs predominantly attenuate the acoustic wave. 
Resonant absorption is suppressed at either high temperature where $P_e = P_g$, or high intensities where $n_{\bf q \eta} \gg 1$ and $P_e \to P_g$. However, the physics of the latter case can only be elucidated by working with the full dynamics governed by the Bloch equations. 

\subsection{Strong fields}

At high acoustic intensities perturbation theory is no longer adequate to calculate the phonon decay rate. 
In this regime one must solve the coupled Heisenberg equations of motion for the TLSs and the acoustic field given by
\begin{align}
\label{}
  \dot{\sigma}_{z,j}  &  = \frac{2}{\hbar} M_j \cdot \xi({\bf r}_j) \sigma_{y,j} \\
   \dot{\sigma}_{y,j}  &  = \frac{1}{\hbar}(E_j + D_j \cdot \xi({\bf r}_j))\sigma_{x,j}  - \frac{2}{\hbar} M_j\cdot\xi({\bf r}_j) \sigma_{z,j}  \\ 
    \dot{\sigma}_{x,j}  & =  - \frac{1}{\hbar}(E_j + D_j\cdot\xi({\bf r}_j)) \sigma_{y,j} \\
  \dot{b}_{{\bf q}\eta} & = -i \Omega_{\bf q \eta} b_{\bf q \eta} 
-  \frac{1}{2}\sum_j g_{{\bf q\eta},j} ( \Delta_j \sigma_{z,j} + 2 \Delta_{0,j}\sigma_{x,j} )
   \end{align}
 where $g_{{\bf q}\eta,j} \equiv \frac{1}{\hbar} \frac{\gamma_\eta}{E_j} q \sqrt{\frac{\hbar}{2 \Omega_{\bf q\eta} \rho V}} e^{-i {\bf q}\cdot {\bf r}_j}$. 

When the acoustic field is driven strongly at angular frequency $\omega \approx \Omega_{\bf q \eta}$ the physics can be greatly simplified. 
Under these conditions we focus our attention solely on the classical steady-state dynamics of the ${\bf q \eta}$-mode. We account for the remaining, thermally populated, phonon
modes with phenomenological damping terms and `Langevin' forces that influence the TLS dynamics. Strong driving also allows the use of the rotating wave approximation (RWA) that we implement by decomposing the strain field and the $x$ and $y$ Pauli operators into positive and negative frequencies i.e. $\xi = \xi^{(+)} + \xi^{(-)}$ where $\xi^{(\pm)} \propto e^{\mp i \omega t}$, and similarly $\sigma_{x,j} = \sigma^{(+)}_{x,j} +\sigma^{(-)}_{x,j}$ where $\sigma^{(\pm)}_{x,j} \propto e^{\mp i \omega t}$. The operator $\sigma_{y,j}$ is similarly decomposed and $\sigma_{z,j}$ is assumed to be time-independent in the steady-state limit.    

After these simplifications the coupled equations of motion for the driven phonon amplitude $\beta_{\bf q \eta}$, and the mean values for the TLS operators, $\langle \sigma_{i,j} \rangle \equiv S_{i,j}$, reduce to 
\begin{align}
\label{EqnsOfMotion}
 &  \dot{S}_{z,j}   = -\frac{1}{T_1}(  S_{z,j}  - w_{0,j}) 
 + \frac{2}{\hbar} M_{\eta,j} \bigg[  \xi_{\bf q \eta}^{(-)}({\bf r}_j)  S^{(+)}_{y,j}   + H.c. \bigg] \\
&   \dot{S}^{(\pm)}_{y,j}     = - \frac{1}{T_2} S^{(\pm)}_{y,j}+\frac{1}{\hbar}E_j S^{(\pm)}_{x,j}  - \frac{2}{\hbar} M_{\eta,j}\xi_{\bf q \eta}^{(\pm)}({\bf r}_j) S_{z,j}  \nonumber \\ 
  &  \dot{S}^{(\pm)}_{x,j}   =  - \frac{1}{T_2} S^{(\pm)}_{x,j}- \frac{1}{\hbar}E_j S^{(\pm)}_{y,j} \nonumber \\
&\dot{\beta}_{{\bf q}\eta} = - i\Omega_{\bf q \eta}\beta_{{\bf q}\eta} + F_{{\bf q}\eta}- \sum_j \Delta_{0,j} g_{{\bf q\eta},j} S^{(+)}_{x,j} \nonumber 
\end{align}
where the TLS phenomenological decay rates $1/T_1$, quantifying the upper state lifetime, and $1/T_2$, characterizing the dephasing rate, arise from the interaction  with the thermal phonon field, and additionally in the latter from spectral diffusion of a given TLS's oscillation frequency $\frac{1}{\hbar}(E_j - D_j\cdot\xi({\bf r}_j))$ as the static background strain field is modified by spin flips of neighboring TLSs \cite{Graebner86}. The strain amplitude of the ${\bf q \eta}$-mode $\xi_{\bf q \eta}^{(+)}$ is given by the coefficient of $b_{\bf q \eta}$ in Eq. \ref{strain} with $b_{\bf q \eta}$ replaced by $\beta_{\bf q \eta}$, $M_{\eta,j}$ is $\gamma_\eta(\Delta_j/E_j)$, and $F_{\bf q \eta}$ is the magnitude of an external drive at angular frequency $\omega$. 
The thermal equilibrium value of $S_{z,j}$ given by $w_{0,j} \equiv - \tanh(E_j/2 k_B T)$ plays the role of a `Langevin' force in this system of equations by ensuring the  return to thermal equilibrium in the absence of driving. 

In steady-state and noting that $\xi^{(+)} = \xi^{(-)\dag}$ the equation for $S_{z,j}$  can be solved in terms of $\xi^{(+)}_{\bf q \eta}$ giving
\begin{equation}
\label{NEqSS}
 S_{z,j}  =  \frac{w_{0,j}}{1+\frac{4 M_{\eta,j}^2 T_1 T_2}{\hbar^2} |\xi_{\bf q \eta}^{(+)}|^2\bigg( \frac{1}{1+\delta_j^2 T_2^2} +\frac{1}{1+\Sigma_j^2 T_2^2}\bigg)  }
\end{equation}
where $\delta_j \equiv E_j/\hbar - \omega$ and $\Sigma_j \equiv E_j/\hbar +\omega$. Eq. \ref{NEqSS} is the nonequilibrium steady-state value for the population inversion at energy $E_j$ under driving by an acoustic beam at angular frequency $\omega$. Eq. \ref{NEqSS} can be used to find the steady-state solution for $S_{x,j}$ which can then be plugged into the equation of motion for the phonon annihilation operator to give the TLS-influenced phonon dynamics 
\begin{align}
\label{ }
& i(\Omega_{\bf q \eta}  -\omega)  \beta_{{\bf q}\eta} \approx  F_{{\bf q}\eta}  \nonumber \\
& +
\underbrace{
 \sum_j   \frac{M^2_{\eta,j} q^2 T_1}{2\hbar \Omega_{\bf q\eta} \rho V} \bigg[
 \frac{1}{1+i T_2 \delta_j} 
  -  \frac{1}{1-iT_2 \Sigma_j}
 \bigg] S_{z,j} 
 }_{- i\Delta \omega_{\bf q\eta} - \frac{1}{2}\Gamma^{\rm res}_{\bf q \eta}} \beta_{{\bf q}\eta}.
\end{align}
The backreaction of the TLSs on the phonons is characterized by a complex susceptibility $- i\Delta \omega_{\bf q\eta} - \frac{1}{2}\Gamma^{\rm res}_{\bf q \eta}$ that modifies the phonon dynamics in two ways: The real part results in dissipation of the phonon beam by resonant absorption, and the imaginary part induces a frequency shift (to be discussed below). 

Assuming the validity of the ergodic hypothesis the sum over $j$ can be converted to an integral.
By assuming that $T_2$ is insensitive to $\Delta_0$, and 
by working in `polar' coordinates $(E,\phi)$, i.e $\Delta = E \cos \phi$ and $\Delta_0 = E \sin \phi$, the dissipation rate can be expressed as 
\begin{align}
\label{ }
\Gamma^{\rm res}_{\bf q \eta} \approx   \frac{P \gamma^2_{\eta} \Omega_{\bf q \eta}}{\hbar \rho v_\eta^2} \int_{-\infty}^\infty d E  \frac{T_2 \tanh(E/2 k_B T)}{
1+(\frac{E}{\hbar} -\omega)^2 T_2^2+\frac{4 \gamma_{\eta}^2 \tau_{\rm min} T_2}{\hbar^2}  |\xi_{\bf q \eta}^{(+)}|^2
},
\end{align}
where $\omega T_2 \gg1$ has been used, and $T_1$ is taken to be approximately given by Eq. \ref{TLSlifetime}.  For $\omega T_2 \gg1$, the integrand is sharply peaked for $E \approx \omega$. Assuming that $T_2$, $\tau_{\rm min}$, and $\tanh(E/2 k_B T)$ vary little over the range $E \in \hbar(\omega-1/T_2,\omega + 1/T_2)$ the integral is given approximately by Eq. \ref{ResAbsorption} where $\Omega_{\bf q \eta} = \Omega$, $\omega = \Omega$, the expression for the acoustic intensity $J = 2 \rho v^3  |\xi_{\bf q \eta}^{(+)}|^2$ has been used, and the critical intensity is actually inversely proportional to $\tau_{\rm min}$ as opposed to $T_1$.
With $f(\Delta,\Delta_0)$ and Eq. \ref{TLSlifetime} the density of TLSs as a function of $E$ and $\tau$ can be derived. This distribution is highly peaked near $\tau_{\rm min}$ (see Eq. 3.8 of \cite{Phillips87}), and thus we approximate $T_1 \gtrsim \tau_{\rm min}$ in Eq. \ref{criticalIntensity}.

\section{Relaxation absorption}

Relaxation absorption results from the modulation of the TLS's energy splitting by a time-dependent strain field. Such an effect was neglected in taking the RWA to arrive at Eqs. \ref{EqnsOfMotion} where we have dropped a $S_{z,j}$-dependent drive term in the equation of motion for the phonon amplitude. The time dependence of $S_{z,j}$ can be properly accounted for, and its effect on the phonon dynamics can be taken into account, leading to dissipation and a frequency shift of the phonons. 

The steady-state amplitude of the time-dependent component of $S_{z,j}$ that dominantly couples to $\beta_{{\bf q}\eta}$ is given by 
\begin{equation}
\label{ }
\delta S^{(+)}_{z,j} = \frac{1}{1- i \omega T_1} \frac{\partial w_{0,j}}{\partial E_j} D_{\eta,j}\xi_{\bf q \eta}^{(+)}({\bf r}_j)
\end{equation}
where $\delta S_{z,j} \equiv S_{z,j} - w_{0,j}$. Accounting for this additional drive term in the equation of motion for the phonons leads to 
\begin{align}
\label{ }
\bigg(i(\Omega_{\bf q \eta}-\omega+ \Delta \omega_{\bf q \eta})+  & \frac{1}{2}\Gamma^{\rm res}_{\bf q \eta} \bigg)\beta_{{\bf q}\eta} =  F_{{\bf q}\eta}  \nonumber \\
&  
 \underbrace{- \frac{i}{V}\sum_j   \frac{ D_{\eta,j}^2  }{ 4 \rho v_\eta^2} \frac{ \Omega_{\bf q \eta} }{1- i\omega T_1} \frac{\partial w_{0,j}}{\partial E_j}}_{- i \Delta \omega^{\rm rel}_{\bf q \eta}-\frac{1}{2} \Gamma^{\rm rel}_{\bf q \eta}}  \beta_{{\bf q}\eta}.
\end{align}
Just as in the case of resonant absorption, the effect of relaxation absorption is quantified by the complex susceptibility $- i \Delta \omega^{\rm rel}_{\bf q \eta}-\frac{1}{2} \Gamma^{\rm rel}_{\bf q \eta}$. Here we focus on the relaxation-induced damping coefficient 
\begin{equation}
\label{ }
\Gamma^{\rm rel}_{\bf q \eta} =  - \frac{2}{V}\sum_j   \frac{D_{\eta,j}^2}{4 \rho v_\eta^2} \frac{ \Omega_{\bf q \eta} \omega T_1}{1+ \omega^2 T_1^2} \frac{\partial w_{0,j}}{\partial E_j}.
\end{equation}
To evaluate $\Gamma^{\rm rel}_{\bf q \eta}$ we employ the ergodic hypothesis and note an important observation about the relative magnitudes of the probing frequency and the inversion decay in our experiment. Assuming that $T_1$ is given by the upper state lifetime of the TLSs then it takes a minimum value calculated by taking $\Delta_0 \to E$ in Eq. \ref{TLSlifetime}.  
Furthermore, the factor $\partial w_0/\partial E$ exponentially suppresses contributions to the integrals from $E$ bigger than $k_B T$. Hence, for the frequencies of interest in our experiment it can be shown that $\omega T_1 \gg 1$ over the integration range contributing to $\Gamma^{\rm rel}_{\bf q \eta}$. Thus, a Taylor expansion of the integrand of $\Gamma^{\rm rel}_{\bf q \eta}$ is justified for small $1/\omega T_1$ and results in 
\begin{equation}
\label{ }
\Gamma^{\rm rel}_{\rm q \eta} =   \frac{P \gamma^2_\eta}{\rho v_\eta^2 k_B T} \int d\Delta \int d \Delta_0 \frac{\Delta^2}{\Delta_0 E^2} \frac{1}{\tau} \frac{1}{\cosh^2\frac{E}{2 k_B T} }
\end{equation}
where we've taken $\Omega_{\rm q \eta}/\omega =1$ and where we've replaced $T_1$ with $\tau$. One arrives at Eq. \ref{RelAbsorption} after performing the integrals.

\end{document}